\documentclass[aps,showpacs,prb,twocolumn,supperscriptaddress]{revtex4-1}
\usepackage{graphicx} 
\usepackage{dcolumn} 
\usepackage{bm} 
\usepackage{graphicx,color,epsfig,rotate}
\usepackage{amsmath,amssymb}
\usepackage{pifont}

\makeatletter
\newcommand{\Rmnum}[1]{\expandafter\@slowromancap\romannumeral #1@}
\makeatother
\newcommand{\dbar}{d\mkern-5mu\mathchar'26}

\begin{document}

\title{Golf-course and funnel energy landscapes: Protein folding concepts in martensites}
\author{N.~Shankaraiah}
\affiliation{TIFR Centre for Interdisciplinary Sciences, 21 Brundavan colony, Narsingi, Hyderabad 500075 , India.}

\begin{abstract}
We use protein folding energy landscape concepts such as golf-course and funnel to study re-equilibration in athermal martensite parameter regime of triangle-to-centered rectangle, square-to-oblique, and triangle-to-oblique transitions under systematic temperature-quench Monte Carlo simulations. On quenching below a transition temperature, the seeded high-symmetry parent-phase austenite that converts to the low-symmetry product-phase martensite, through autocatalytic twinning or elastic photocopying, has both rapid conversions and incubation-delays in the temperature-time-transformation phase diagram. We find the rapid (incubation-delays) conversions at low (high) temperatures arises from the presence of large (small) size of golf-course edge that has funnel inside for negative energy states. In the incubating state, the strain structure factor enters into the Brillouin zone golf-course through searches for finite transitional pathways which closes off at the transition temperature with Vogel-Fulcher divergences that are insensitive to Hamiltonian energy scales and log-normal distributions, as signatures of dominant entropy barriers. The crossing of the entropy barrier is identified through energy occupancy distributions, Monte Carlo acceptance fractions, heat emission and internal work. The above ideas had previously been presented for the scalar order parameter case. Here we show similar results are also obtained for vector order parameters. 
\end{abstract}


\maketitle

\vskip 0.5truecm

\section{Introduction}
\
Energy landscape concepts \cite{R1} such as golf-course and funnel are used in proteins \cite{R2,R3,R4} to understand the folding kinetics, in temperature (on $X$-axis) and time (on $Y$-axis) diagram \cite{R5,R6}: (i) rapid folding below a transition temperature and slow folding above it; and (ii) U-shaped folding curves. The rapid folding change to become slow at low temperatures on changing the roughness of funnel \cite{R7}.  
The glassy ruggedness and slope of the folding funnel are estimated \cite{R3} from experimental data. 
In a simple model of brownian particle searching outside a golf-course (``unfolded state'') for a funnel inside it (''folded state'') find entropic barriers at the golf-course edge and exponential relaxation kinetics \cite{R8}.   
In an off-lattice Go-model for inherent structure energy landscape of proteins a time-dependent effective temperature is obtained from internal energy and entropy \cite{R9}. 
In a topology-based dynamical model, the Vogel-Fulcher divergences that are well known in glasses \cite{R10} and broad distributions are found for unfolding of proteins \cite{R11}.
Such slow relaxations in glasses are understood to arise from entropy barriers {\it alone} in a simple microscopic model without energy barriers \cite{R12,R13}.

Martensites are materials \cite{R14,R15} that undergo a diffusionless and displacive first-order phase transition on cooling or under external stress, from high-temperature high-symmetry parent austenite unit-cell to low-temperature low-symmetry product martensite unit-cells or variants ($N_v$). Steels, shape memory alloys, high-$T_c$ superconductors, ceramics, oxides and proteins are few examples \cite{R14,R15}. A subset of physical strain components ($N_{OP}$) are the order parameters (OP) and the remaining non-OP strains are minimized subject to a no-defect Saint-Venant compatibility constraint that induces scale-free, power-law, anisotropic interactions which orients the domain walls in preferred crystallographic directions \cite{R16}.
Martensites can have exponentially large number of multivariant twinned states or nonuniform metastable local minima competiting with a single uniform global minimum \cite{R17}. 
Martensites are classified based on conversion times \cite{R18,R19}: (i) {\it athermal}, which are expected to have rapid austenite-to-martensite conversions in milli-seconds on quenching below a transition temperature and no conversions above it; and (ii) {\it isothermal} can have slow conversions in minutes or hours. 

Systematic temperature-quench Monte Carlo (MC) simulations are performed on strain-pseudospin clock-zero model Hamiltonians in 2-spatial dimensions for scalar-OP $(N_{OP}=1)$ square-rectangle $(SR, N_{v}+1=3)$ transition \cite{R17,R20}, and vector-OP $(N_{OP}=2)$ triangle-centered rectangle $(TCR, N_{v}+1=4)$, square-oblique $(SO, N_{v}+1=5)$, and triangle-oblique $(TO, N_{v}+1=7)$ transitions \cite{R21} and found both isothermal and athermal martensite parameter regimes. The pseudospin strain textures obtained from MC simulations and local meanfield \cite{R17,R20,R21,R22} are in very good agreement with experiments \cite{R23,R24,R25}.
In the {\it temperature-time-transformation} (TTT) diagram with temperature on $X$-axis and time on $Y$-axis \cite{R17,R20,R21}: (i) {\it athermal} martensites have rapid conversions below a transition temperature and delays above it as in experiments, with Vogel-Fulcher divergences that are insensitive to Hamiltonian energy scales, understood from the presence of non-activated {\it entropy barriers}; and (ii) {\it isothermal} ones have U-shaped conversion curves, as expected to arise from activated {\it energy barriers}.  
The shape of TTT curves transform from rapid to slow (or athermal to isothermal) at low temperatures on changing the material elastic stiffness constant \cite{R17,R20,R21}.


In the athermal martensite regime, golf-course and funnel energy landscapes that appear in {\it Fourier space}  naturally in a simple 3-state strain-pseudospin clock-zero model Hamiltonian for scalar-OP $(N_{OP}=1)$ square-rectangle transition are used to study the rapid and slow austenite-to-martensite conversions and re-equilibration under systematic temperature-quench MC simulations \cite{R26}.
Energy landscape concepts in martensites are used in other contexts \cite{R27,R28,R29,R30,R31,R32}.
In perovskite manganites, the strain-induced metal-insulator phase coexistence is understood through an elastic energy landscape \cite{R27}. 
In a binary alloy system, the crystallization of strain glass and it's properties are studied using frustrated free-energy landscape \cite{R28}. 
In iron nanoislands system, an energy landscape is modeled to study the dynamics of electrically driven body-centered cubic to face-centered cubic phase transition \cite{R29}.
In a single crystal, Peierls-Nabarro energy landscape is used to model cubic-monoclinic transition \cite{R30}.
In a shape memory alloy, the distortion-shuffle energy landscape is used to identify the energy barrier for cubic-orthorhombic transition \cite{R31}.
The global complex energy landscapes are proposed to model elastic moduli and energy barriers for cubic-tetragonal and cubic-monoclinic transitions \cite{R32}.


In this paper, the athermal regime re-equilibration and nature of entropy barriers are studied in 4-state, 5-state, and 7-state strain-pseudospin clock-zero model Hamiltonians for vector-OP $(N_{OP}=2)$ triangle-centered rectangle (TCR), square-oblique (SO), and triangle-oblique (TO) transitions using naturally appearing {\it Fourier space} golf-course and funnel energy landscapes, and MC acceptance fractions. 
The rapid and slow incubation-delay conversions are found to arise from the presence of large and small size of the golf-course edges. The Vogel-Fulcher conversion-delays that are insensitive to Hamiltonian energy scales are found to have Log-normal distributions that are signatures of rare events \cite{R33}. The number of successful conversions, that are also insensitive to energy scales, vanishes where the entropy barriers diverges \cite{R12,R17,R20}.
In the incubating state, the crossing of entropy barrier is identified in energy occupancy distributions, MC acceptance fractions and heat and work releases as the structure factor enters into the Brillouin zone (BZ) golf-course through searches for rare energy-lowering pathways and {\it elastic photocopying} \cite{R34,R35}.

This paper is organized as follows. In sec.II, we discuss the strain-pseudospin clock-zero model Hamiltonians and the MC simulation techniques. Sec.III contains golf courses, funnels, and conversion times; evolution of strain textures in coordinate and Fourier spaces; and energy occupancy distributions of structure factor. We also present the MC acceptance fractions and work and heat releases.
Finally, sec. IV is a summary with an overview of potential further work.

\section{Strain-pseudospin clock-zero model Hamiltonians}

The pseudospin clock-zero model Hamiltonians for TCR, SO and TO transitions were systematically derived from scaled continuous-strain free-energies \cite{R16}. We outline here for completeness. 
In $d$-spatial dimensions, the distortions of a unit-cell are described by $\frac{1}{2}d(d+1)$ Cartesian strain tensor components $e_{\mu \nu}$ and physical strains $e_{\alpha}$ are linear combinations of these components ($e_{\mu\nu}$). In $d=1$ dimensions, there is only one strain $e=\partial u(x)/\partial x$. In $d=2$ dimensions, there are three distinct physical strains namely dilatational or compressional $(e_1)$, rectangular or deviatoric $(e_2)$ and shear $(e_3)$,
$$ e_1=\frac{1}{\sqrt{2}}(e_{xx}+e_{yy}), e_2=\frac{1}{\sqrt{2}}(e_{xx}-e_{yy}), e_3=e_{xy}, ~~~~~~~~~(2.1)$$
where $e_{xy},e_{yx}$ are tilts and $e_{xx},e_{yy}$ are stretches or compressions along $x$ and $y$ directions of a unit cell. 
A subset of physical strains ($N_{op}$) are the OP and the remaining are non-OP strains, which cannot be set to zero. In d-dimensions, $\frac{1}{2}d(d+1)-N_{op}$ are the non-OP strains that are minimized subject to $\frac{1}{2}d(d-1)$ Saint-Venant compatibility constraints that says all the distorted unit cells fit together smoothly so that no dislocations generated throughout the system. For TCR, SO, and TO transitions, we have $\vec e=(e_2,e_3)$ as two-component vector-OP $(N_{OP}=2)$ and $e_1$ as non-OP strain inducing single compatibility constraint.       

The scaled free energy \cite{R16} has a transition specific Landau term $\bar F_{L}$ that has ($N_{v}+1$) degenerate energy minima at the first-order transition; a Ginzburg term for domain-wall energy costs, quadratic in the OP gradients ${\bar F}_G $; and a compatibility-induced term harmonic in the non-OP strains ${\bar F}_{non}$. Thus 
 $$F = E_0 [ {\bar F}_L + {\bar F}_G + {\bar F}_{non} ],~~~~~~~~ (2.2)$$
where $E_0$ is an elastic energy per unit cell.
 
The discrete-strain pseudospin model Hamiltonians are derived \cite{R16} by choosing continuous-strain OP $\vec e= (e_2, e_3)$ values only at the $N_{v}+1$ Landau minima $\vec e (\vec r) = |e| (\cos\phi , \sin\phi) \rightarrow  {\bar \varepsilon }(\tau) \vec S(\vec r)$ into the total free energy of (2.2),
 $$\beta H( \vec S) \equiv \beta { F} (\vec e \rightarrow {\bar \varepsilon } \vec S ).~~~~~~~ (2.3)$$

The Landau term in Fourier space becomes,
$$ H_L  (\vec S)=  {\bar \varepsilon}^2 \sum_{\vec r} g_L (\tau)  {\vec S}^2(\vec r)= {\bar \varepsilon}^2 \sum_{\vec k} g_L (\tau) |\vec S(\vec k)|^2, ~~(2.4a)$$
where $g_L = \tau - 1  + ( \bar \varepsilon - 1)^2$ with ${\bar \varepsilon}^2 (\tau) =\frac{3}{4} \{ 1 + \sqrt{ 1 - 8 \tau /9} \}$ for TCR; and $g_L = \tau - 1  + ( {\bar \varepsilon}^2 - 1)^2$ with ${\bar \varepsilon}^2 (\tau) =\frac{2}{3} \{ 1 + \sqrt{ 1 - 3 \tau /4} \}$ for SO and TO transitions.
The scaled temperature is defined as $$ \tau=\frac{T-T_c}{T_0-T_c},~~ (2.4b)$$ where $T_0$ is the first-order Landau transition temperature and $T_c$ is the metastable austenite spinodal temperature.

The Ginzburg term becomes,
$$ H_G (\vec \nabla \vec S) = \xi^2  {\bar \varepsilon}^2 \sum_{\vec r} (\vec \nabla {\vec S})^2  = \xi^2 {\bar \varepsilon}^2 \sum_{\vec k} {\vec K}^2 |{\vec S}(\vec k)|^2~~~(2.5), $$
where $\xi$ is the domain-wall thickness constant.

\begin{figure}[ht]
 \begin{center}
 \includegraphics[height=2.5cm, width=9.0cm]{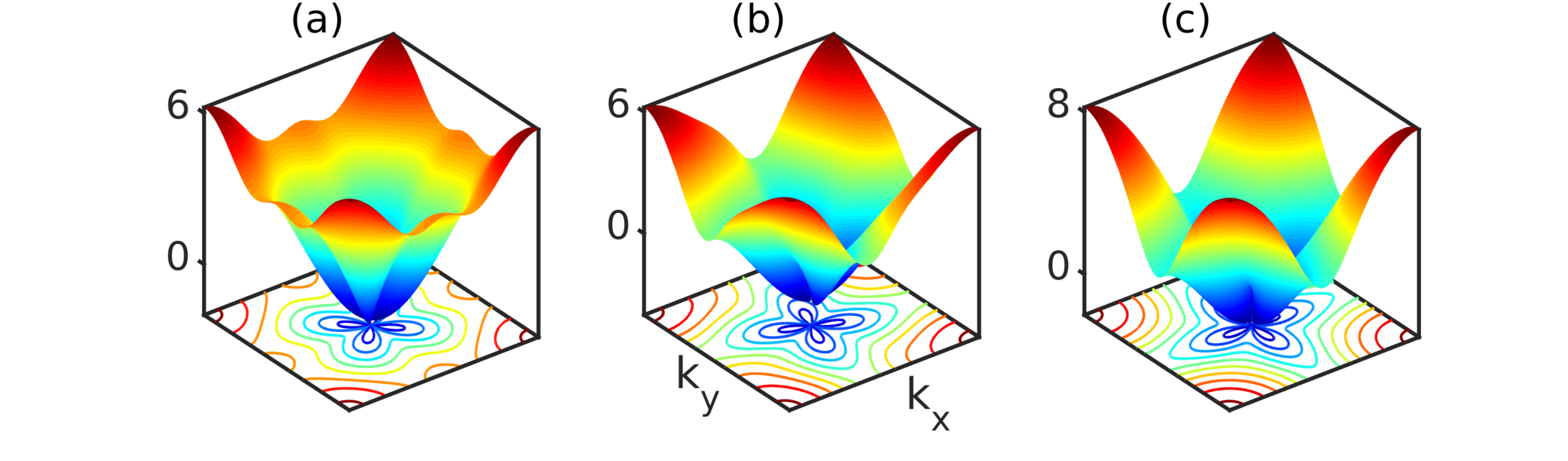}
 \caption{(Color online){\it The martensite energy landscape spectrum:} The relief plot of $\epsilon_{\ell\ell^{'}}(\vec k)$, dimensionless martensite energy spectrum (a) $\epsilon_{22}(\vec k)$; (b) $\epsilon_{23}(\vec k)$; and (c) $\epsilon_{33}(\vec k)$ for TCR, SO, and TO transitions.}
 \end{center}
 \label{Fig.1}
 \end{figure} 
 
The harmonic non-OP term is minimized subject to Saint-Venant compatibility constraint $\vec \nabla \times [ \vec \nabla \times \vec e(\vec r) ]^{T}=0$ for physical strains \cite{R16} that guarantees the lattice integrity during distortion of the unit cells throughout the system,

 $$ \vec \Delta ^2 e_1 - (\Delta_x ^2 - \Delta_y ^2) e_2 - 2\Delta_x \Delta_y  e_3 = 0;~~(2.6a)$$ 
with gradient terms as difference operators ${\vec \nabla} \rightarrow {\vec \Delta}$ for sites $\vec r$ on a computational grid. 
In Fourier space, with $\mu=x,y$ and $k_\mu  \rightarrow K_\mu (\vec k) \equiv 2 \sin (k_\mu /2)$, equation (2.6a) becomes,
$$ O_1 e_1 + O_2 e_2 + O_3 e_3 =0, ~~ (2.6b)$$
 with the coefficients $O_1=-\frac{1}{\sqrt{2}}{\vec K } ^2$, $O_2=\frac{1}{\sqrt{2}}({ K_x ^2 - K_y ^2 }) $, and $O_3={2K_x  K_y }$ for square lattice; and $O_1=-{\vec K } ^2 $, $O_2=({ K_x ^2 - K_y ^2 })$, and $O_3={2K_x  K_y }$ for triangular lattice. Here, $\vec K^2=(K^2_x+K^2_y)$.

The minimization of non-OP strain generates scale-free and power-law anisotropic interactions between the OP strains and becomes,
$$ H_{C} = \frac{A_1}{2} {\bar \varepsilon}^2 \sum_{\vec k} \sum_{\ell, \ell^{'}=2, 3} S_{\ell}(\vec k) ~U_{\ell \ell^{'}}(\vec k) ~S^{\ast}_{\ell^{'}}(\vec k) ~(2.7)$$
where $A_1$ is {\it elastic stiffness} constant. The kernels plotted in \cite{R21} are $ U_{22} (\vec k) =  \nu (O_2/O_1) ^2$ , $U_{23} (\vec k) =  \nu (O_2 O_3/O_1) ^2$ , $U_{33} (\vec k) =  \nu  (O_3/O_1) ^2$ with $\nu=(1-\delta_{\vec k,0})$. 

The Hamiltonian is diagonal in Fourier space \cite{R21}, 

$$H (\vec S)= \frac{K_0}{2} \sum_{\vec k} \sum_{\ell,\ell^{'}=2,3}  \epsilon_{\ell \ell^{'}} (\vec k) S_{\ell} (\vec k) {S^{\ast}_{\ell^{'}}} (\vec k).~~(2.8a)$$ 
The dimensionless martensite strain spectrum,
$$\epsilon_{\ell \ell^{'}} (\vec k) \equiv K_0 [\{ g_L(\tau) +  \xi^2 { \vec K}^2  \}\delta_{\ell \ell^{'}}  + \frac{A_1}{2} (1-\delta_{\vec k,0})U_{\ell \ell^{'}} ({\vec k})],~(2.8b)$$
is plotted in Fig.1 for $T=0.79$, which depicts the energy landscapes similar to that used in protein folding \cite{R2,R3,R4,R5,R6,R7,R8}. Here, $K_0(T)=2E_0 {\bar \varepsilon(T)}^2$. 
This is a clock-zero model Hamiltonian with a austenite $\vec S=(S_2,S_3)=(0,0)$ and $N_v$ martensite variants: 
$$\vec S= (1 , 0) , (-\frac{1}{2} , \pm \frac{\sqrt 3}{2});   (\pm \frac{1}{2} , \pm \frac{1}{2}); (\pm 1,0), (\pm \frac{1}{2}, \pm \frac{\sqrt 3}{2})~(2.8c) $$
for TCR  ($N_{v}+1=4$), SO ($N_{v}+1=5$), and TO ($N_{v}+1=7$) transitions respectively.

 \begin{figure}[ht]
 \begin{center}
 \includegraphics[height=6.0cm, width=8.0cm]{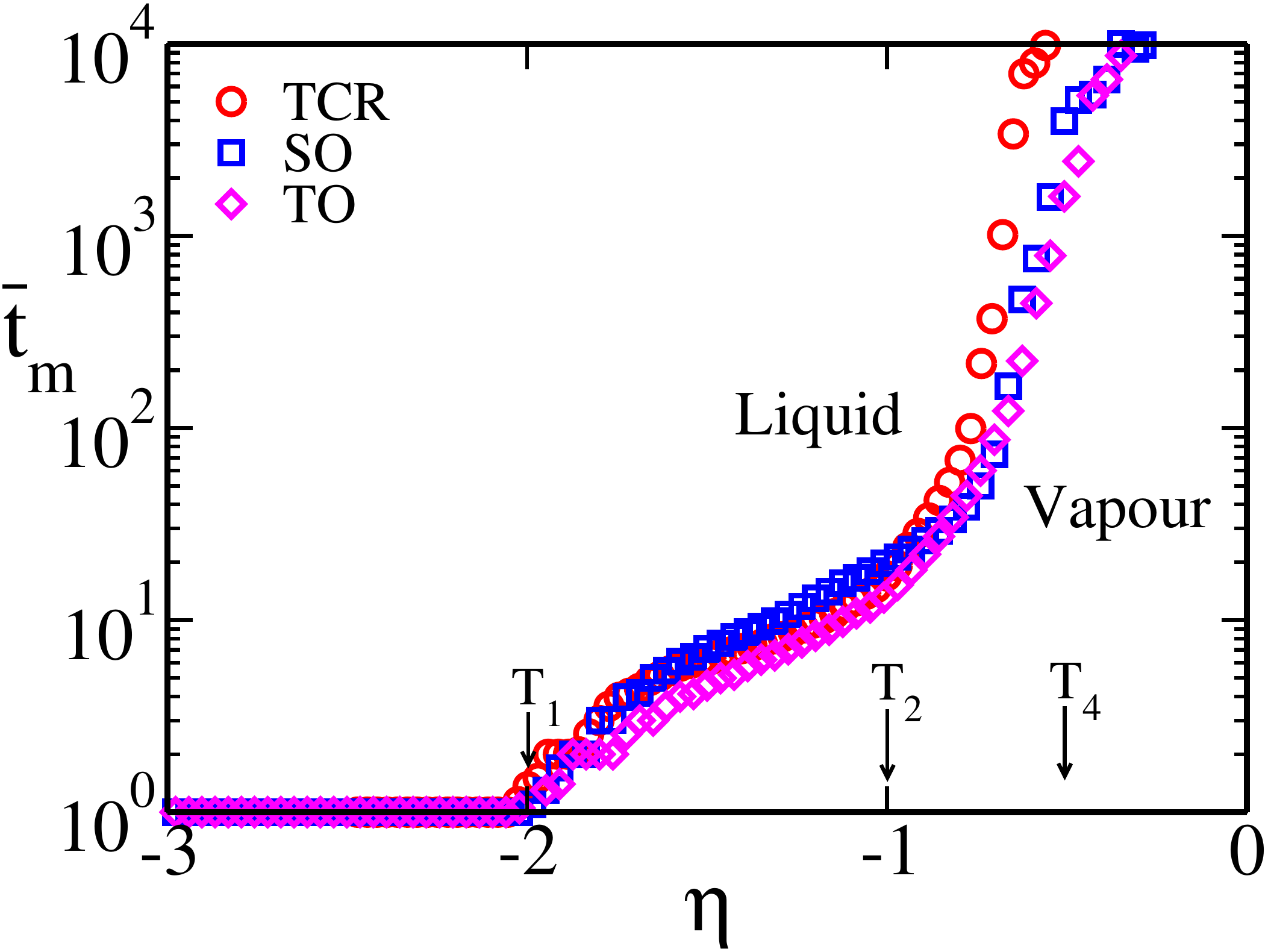}
 \caption{(Color online) {\it Temperature-time-transformation phase diagram:} On quenching, conversion times $\bar t_m$ for domain wall vapour to liquid have rapid, moderate and delayed conversions at the transition temperatures $T=T_1 (=0.15,0.38,0.38)$, $T=T_2 (=0.47,0.55,0.61)$, and  $T=T_4 (=0.68,0.81,0.81)$ that are marked in the scaled temperature variable $\eta(T)$ for TCR, SO, and TO transitions.}
 \end{center}
 \label{Fig.2}
 \end{figure}

  \begin{figure}[ht]
 \begin{center}
 \includegraphics[height=5.8cm, width=8.0cm]{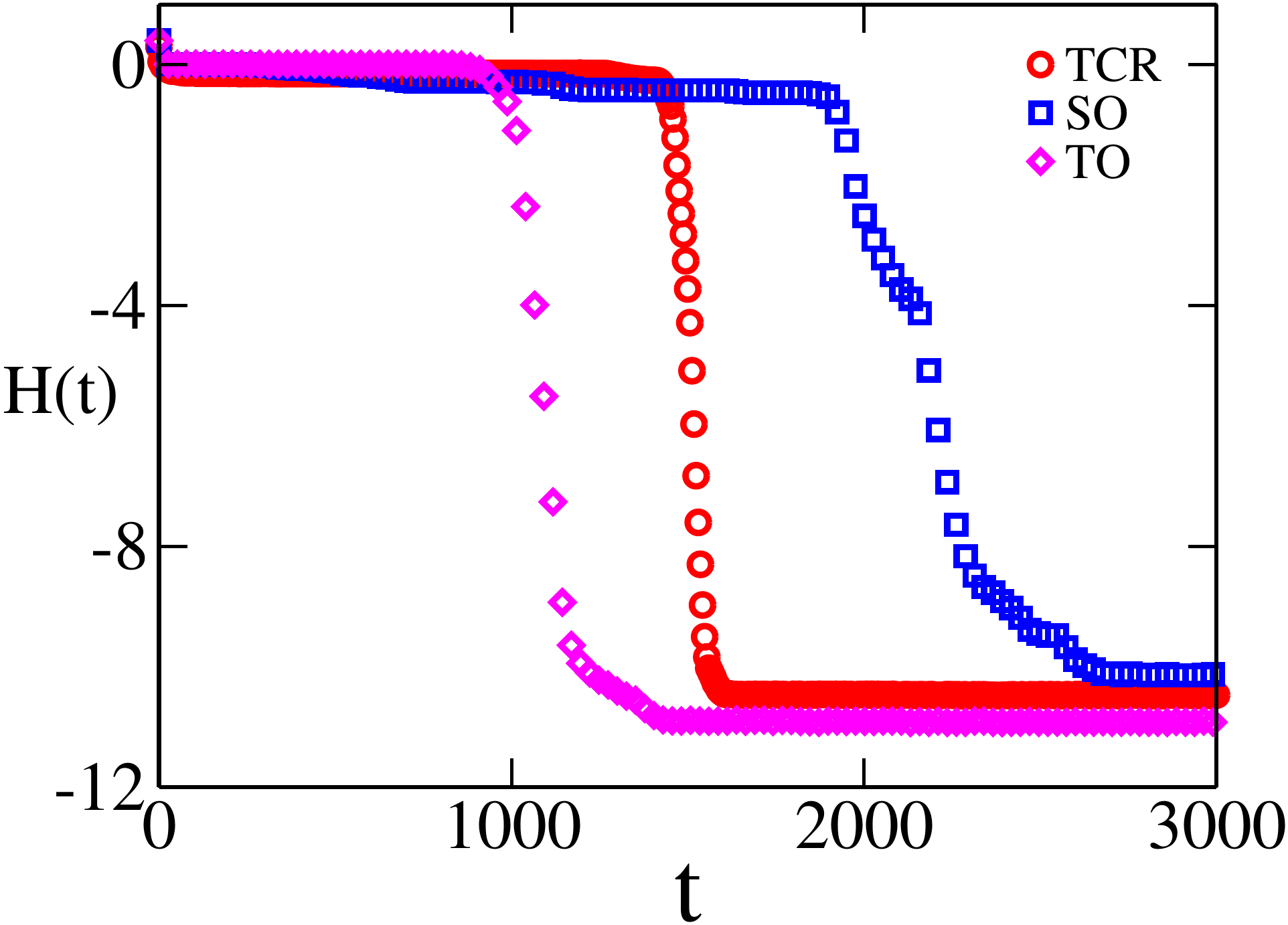}
 \caption{(Color online) {\it Evolution of Hamiltonian energy:} On quenching to a temperature $T \simeq T_4$, the total Hamiltonian energy $H(t)$ versus time $t$ (in MCS) showing incubation at constant energy $H(t) \sim 0$ in TCR, SO, and TO transitions.}
 \end{center}
 \label{Fig.3}
 \end{figure}
 
We have carried out systematic MC temperature quench and hold simulations on a square lattice in 2D \cite{R21}. We quench the austenite with $2\%$ of randomly sprinkled martensite seeds of $N_{v}$ strain-pseudospin values, at $t=0$, 
\begin{figure}[ht]
\begin{center}
\includegraphics[height=2.5cm, width=8.5cm,trim=2.5cm 0.0cm 2.0cm 0.0cm,clip]{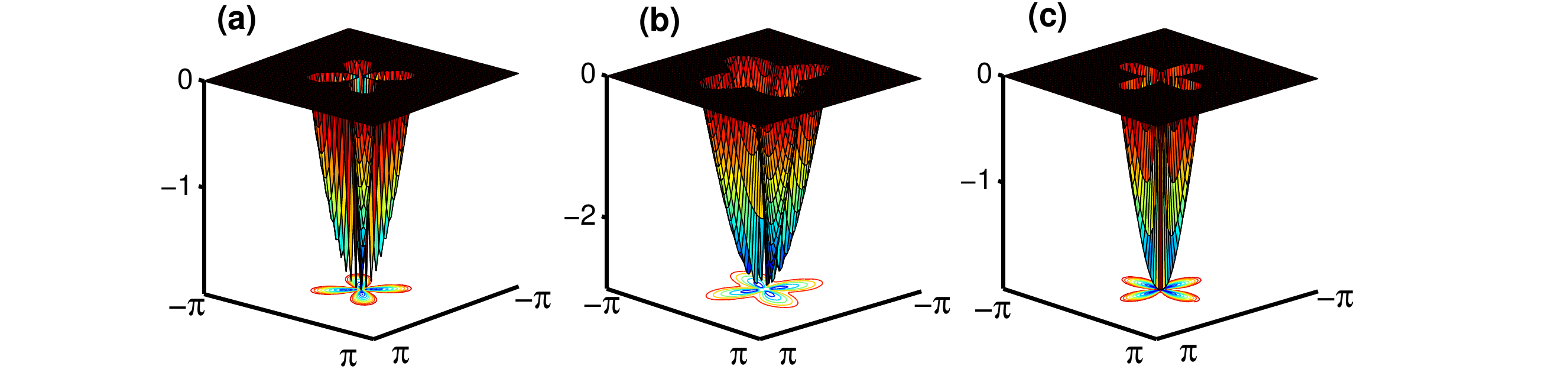}\\
\includegraphics[height=2.2cm, width=8.0cm,trim=3.4cm 0.0cm 0.5cm 0.0cm,clip]{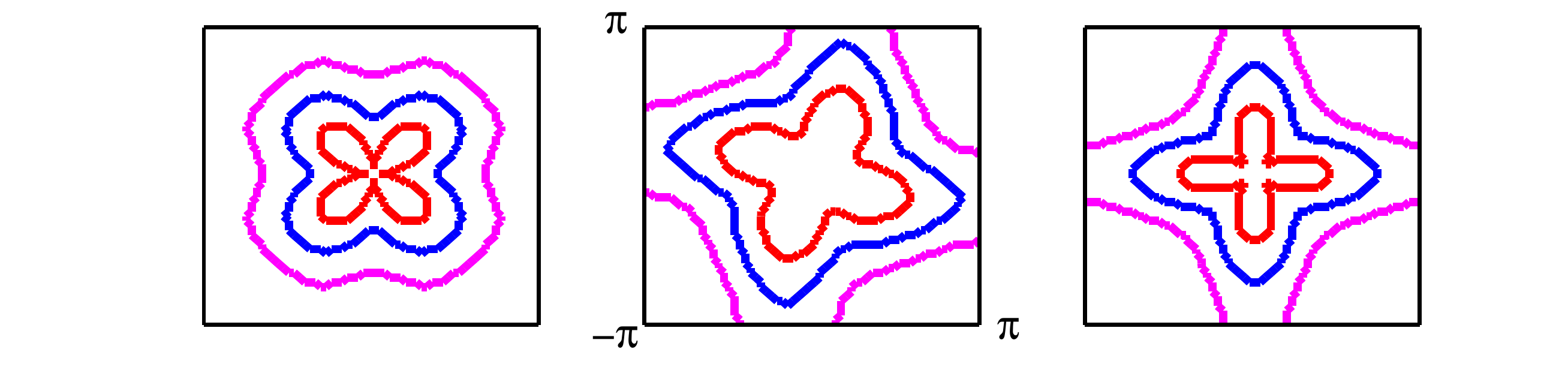}
\caption{(Color online) {\it Anisotropic golf-courses and funnels:} Relief plot of relavent martensite spectrum $\epsilon_{\ell \ell^{'}}(\vec k)$ versus $\vec k$ in Brillouin zone for $T=0.79 (< T_4)$ showing anisotropic golf-course, with a zero-energy plane that has funnel for negative energies as shown in (a) $\epsilon_{22}(\vec k)$, (b) $\epsilon_{23}(\vec k)$, and (c) $\epsilon_{33}(\vec k)$. The edge of golf-course $\epsilon_{22}(\vec k)=0$, $\epsilon_{23}(\vec k)=0$, and $\epsilon_{33}(\vec k)=0$ is plotted respectively, in the bottom row, at $T_1 (=0.38)$ (in pink), $T_2 (=0.55)$ (in blue), and $T_4 (=0.81)$ (in red) of SO transition.}
\end{center}
\label{Fig.4}
\end{figure}
\begin{figure}[ht]
\begin{center}
\includegraphics[height=2.5cm, width=8.5cm]{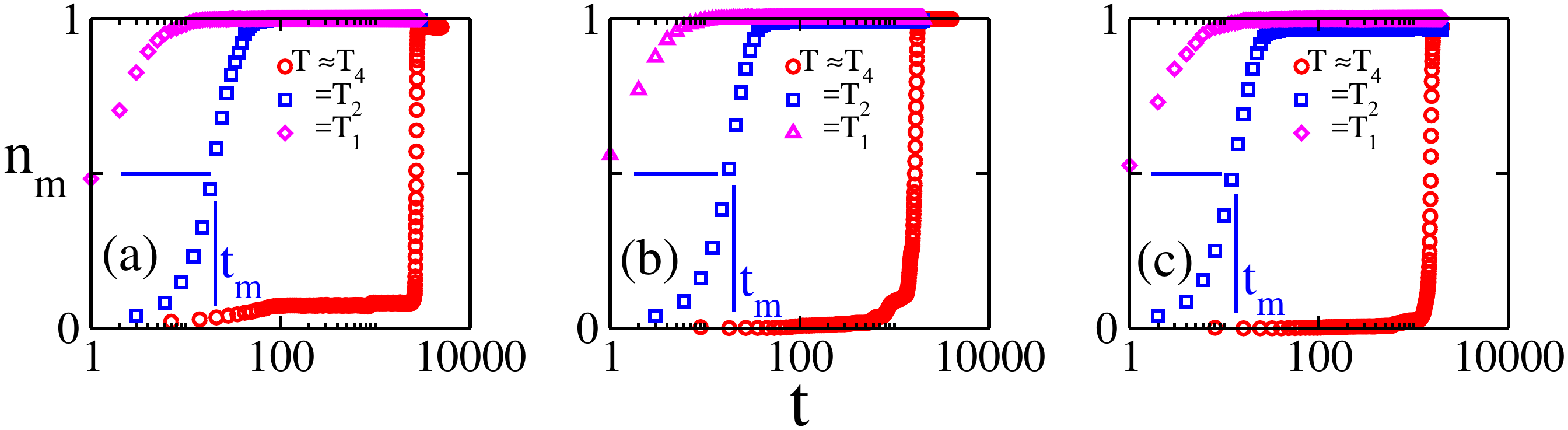} 
\caption{(Color online) {\it Martensite fraction:}  On quenching to temperatures $\Delta T=-0.06~(T \approx T_4), -0.16~(T=T_2), -0.52~(T=T_1)$, martensite fraction $n_m(t)$ versus time $t$ showing incubation-delays and rapid conversions respectively in (a) TCR (b) SO and (c) TO transitions. The conversion time $t=t_m$ is marked at $n_m(t_m)=0.5$.} 
\end{center}
\label{Fig.5}
\end{figure}
\begin{figure}[ht]
\begin{center}
\includegraphics[height=6.0cm, width=8.0cm]{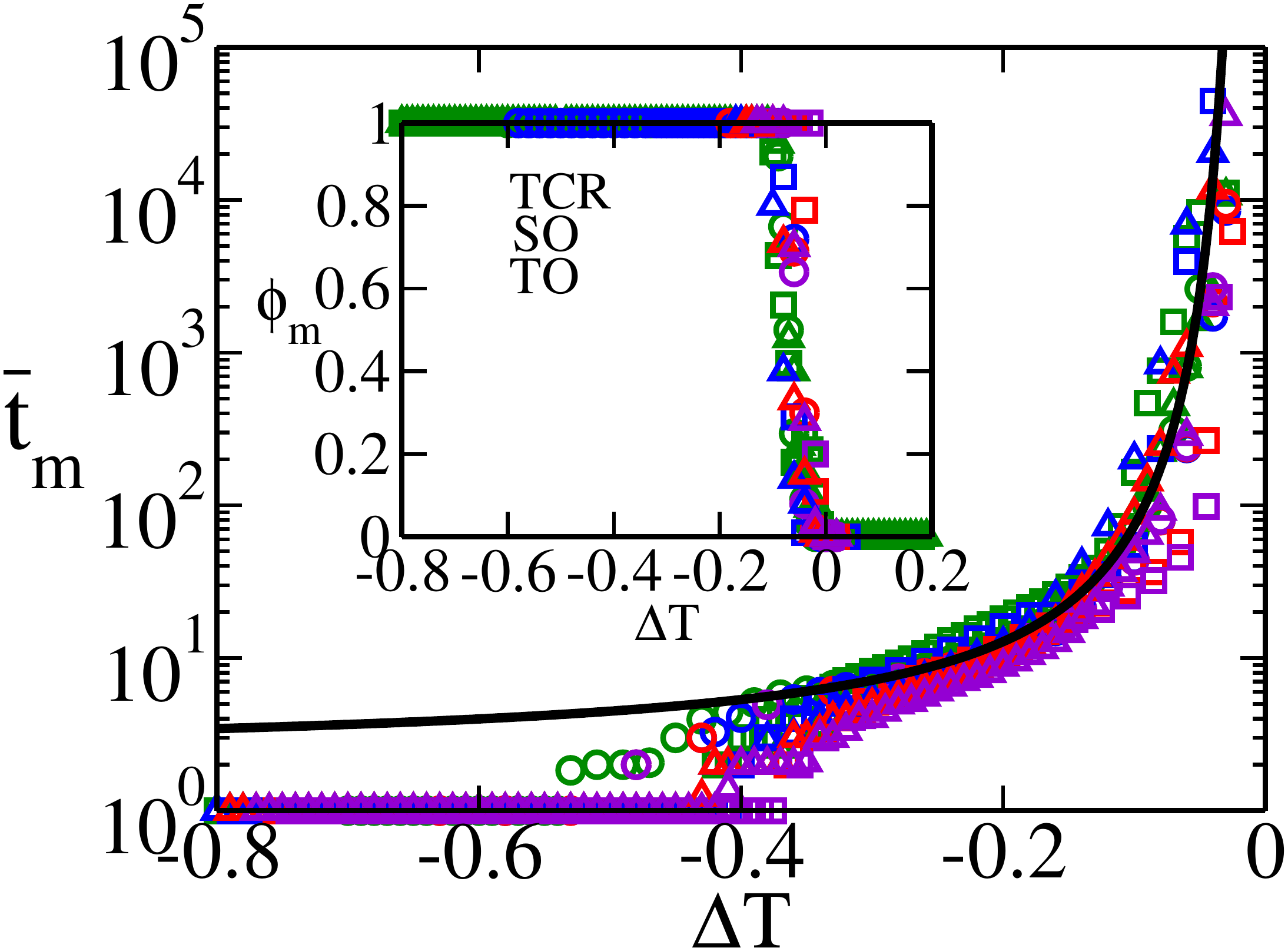} 
\caption{(Color online) {\it Vogel-Fulcher conversions and vanishing pathways:} The Log-linear plot of mean conversion times $\log_{10}(\bar t_m)$ versus temperature deviations $\Delta T=T-T_4$ showing, at transition $T_4$, Vogel-Fulcher divergences that are independent of energy scales $E_0=3,4,5$, and $6$. {\it Inset:} The success fraction $\phi_m$ vs $\Delta T$ showing, at the transition $T_4$, vanishing of rare conversion pathways and insensitivity to $E_0$.} 
\end{center}
\label{Fig.6}
\end{figure}
\begin{figure}[ht]
\begin{center}
\includegraphics[height=2.5cm,width=8.5cm]{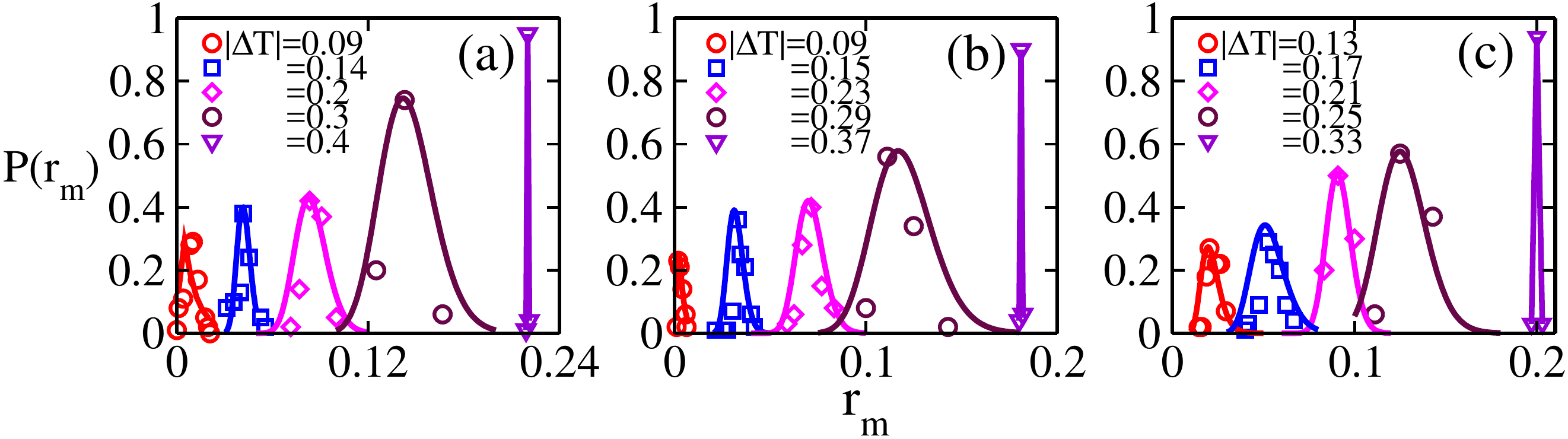}
\caption{(Color online) {\it Log-normal distribution of conversion-rates:} The probability distributions $P(r_m)$ versus conversion rates $r_m$ plotted for different temperature deviations $\Delta T$ for (a) TCR (b) SO and (c) TO transitions. The log-normal distributions (solid lines) are the best fits to the data (symbols).} 
\end{center}
\label{Fig.7}
\end{figure}
to below the Landau transition $T<<T_0$ and held for $t \le t_h$ MC sweeps. The Metropolis algorithm \cite{R36,R37} is used for acceptance of energy changes that are calculated through fast Fourier transforms. We visit all $N=L \times L$ sites randomly, but only once, in each MC sweep (MCS).  Parameters are $L=64, T_0 = 1$; $T_c /T_0 = 0.9$,  $\xi = 1; A_1 =4; 2 A_1 / A_3 = 1$; $E_0 =3, 4, 5, 6; t_h \leq 10,000$, and conversion times are averaged over $N_{runs}=100$ runs.

The TTT phase diagram for TCR, SO and TO transitions is depicted in Fig. 2 as a log-linear plot of conversion times $\bar t_m$ versus scaled temperature variable \cite{R17,R21} $\eta(T)=\{g_L(\tau)+A_1[U]/2\}/2 \xi^2$ that shows boundary between domain-wall (DW) vapour and liquid phases. Here $[U] \simeq 0.5$ is the BZ average of $U_{\ell \ell^{'}}(\vec k)$ in TCR,SO, and TO transitions. The crossover temperatures that are understood through the parametrization of pseudospin strain textures with an effective droplet energy \cite{R21} are $T=T_1$ or $\eta(T)=-2$ where conversion-times (in units of MCS) $\bar t_m \sim 1$, $T=T_2$ or $\eta(T)=-1$ where $\bar t_m \sim 10$ and $T=T_4$ or $\eta(T) \sim -0.5$ where $\bar t_m$ diverges. For $T > T_4$, there are no conversions to martensite and hence the initial seeds disappear to go back to austenite.

To study the re-equilibration under a quench-and-hold protocol, we track the dynamic structure factor \cite{R26}
  $$\rho (\vec k, t) \equiv |\vec S(\vec k, t)|^2 ,~~~(2.9)$$  
and its BZ average, the martensite fraction,  
 $$n_m(t)=\frac{1}{N}\sum_{\vec k} \rho (\vec k, t) = \frac{1}{N} \sum_{\vec k} | \vec S (\vec k, t)|^2,~~(2.10) $$ 
 that is zero in austenite and unity in twinned or uniform martensite.
 We define conversion time $t=t_m$ when $n_m(t_m)=0.5$ or $50\%$ \cite{R17,R20,R21,R26}. On quenching to different temperatures, we find the conversion success-fraction \cite{R17} $\phi_m$ that is the number of successful conversion pathways to martensite out of $N_{runs}$.

 \begin{figure*}[ht]
\begin{center}
\includegraphics[height=8.0cm,width=16.0cm,trim=3.0cm 2.0cm 2.0cm 1.0cm,clip]{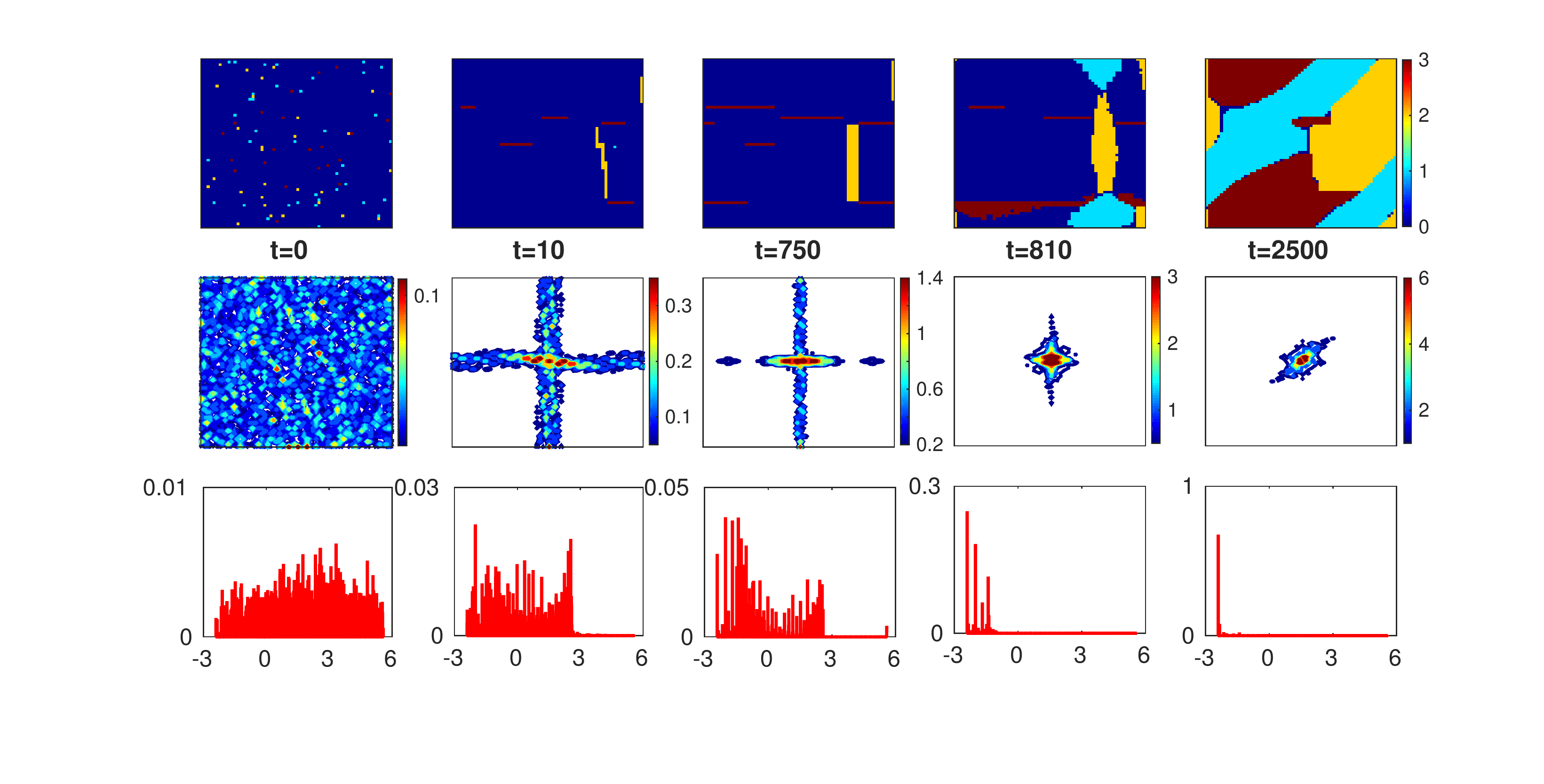}
\caption{(Color online) {\it Evolution of strain textures and energy occupancy in TCR transition:} On quench-and-hold to $T=0.63$, snapshots of OP textures in Brillouin zone as $ln(1+|\vec S(\vec k)|^2)$ contours ({\it second row}) of corresponding coordinate space textures ({\it first row}) showing incubation during domain-wall vapour to liquid which finally converts to crystal. See movies  \cite{R39} of these evolutions in both coordinate and Fourier spaces. The energy occupancy $\rho_{22}(\epsilon,t)$ of structure factor vs $\epsilon_{22}(\vec k)$ ({\it third row}) showing the shifting of density of states when the distribution enters into the golf-course. See text.}
\end{center}
\label{Fig.8}
\end{figure*}

\begin{figure*}[ht]
\begin{center}
\includegraphics[height=8.0cm,width=16.0cm,trim=3.0cm 2.0cm 2.0cm 1.0cm,clip]{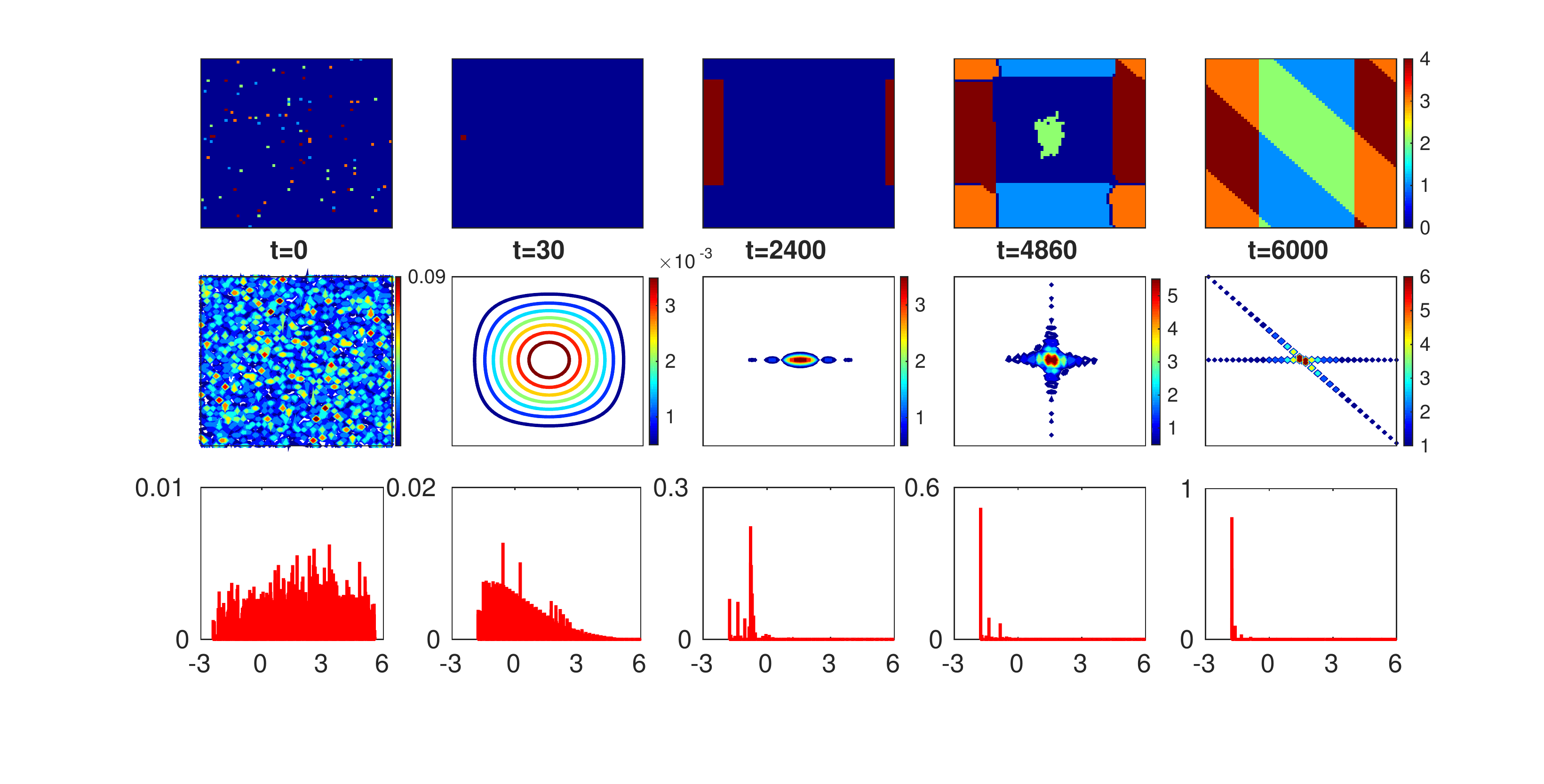}
\caption{(Color online) {\it Evolution of strain textures and energy occupancy in SO transition:} On quench-and-hold to $T=0.79$, snapshots of OP textures in Brillouin zone $ln(1+|\vec S(\vec k)|^2)$ contours ({\it second row}) of corresponding coordinate space textures ({\it first row}) showing incubation during domain-wall vapour to liquid which finally converts to crystal. See movies \cite{R39} of these evolutions in both coordinate and Fourier spaces. The energy occupancy $\rho_{22}(\epsilon,t)$ of structure factor vs $\epsilon_{22}(\vec k)$ ({\it third row}) showing the shifting of density of states when the distribution enters into the golf-course. See text.}
\end{center}
\label{Fig.9}
\end{figure*}

\begin{figure*}[ht]
\begin{center}
\includegraphics[height=8.0cm,width=16.0cm,trim=3.0cm 2.0cm 2.0cm 1.0cm,clip]{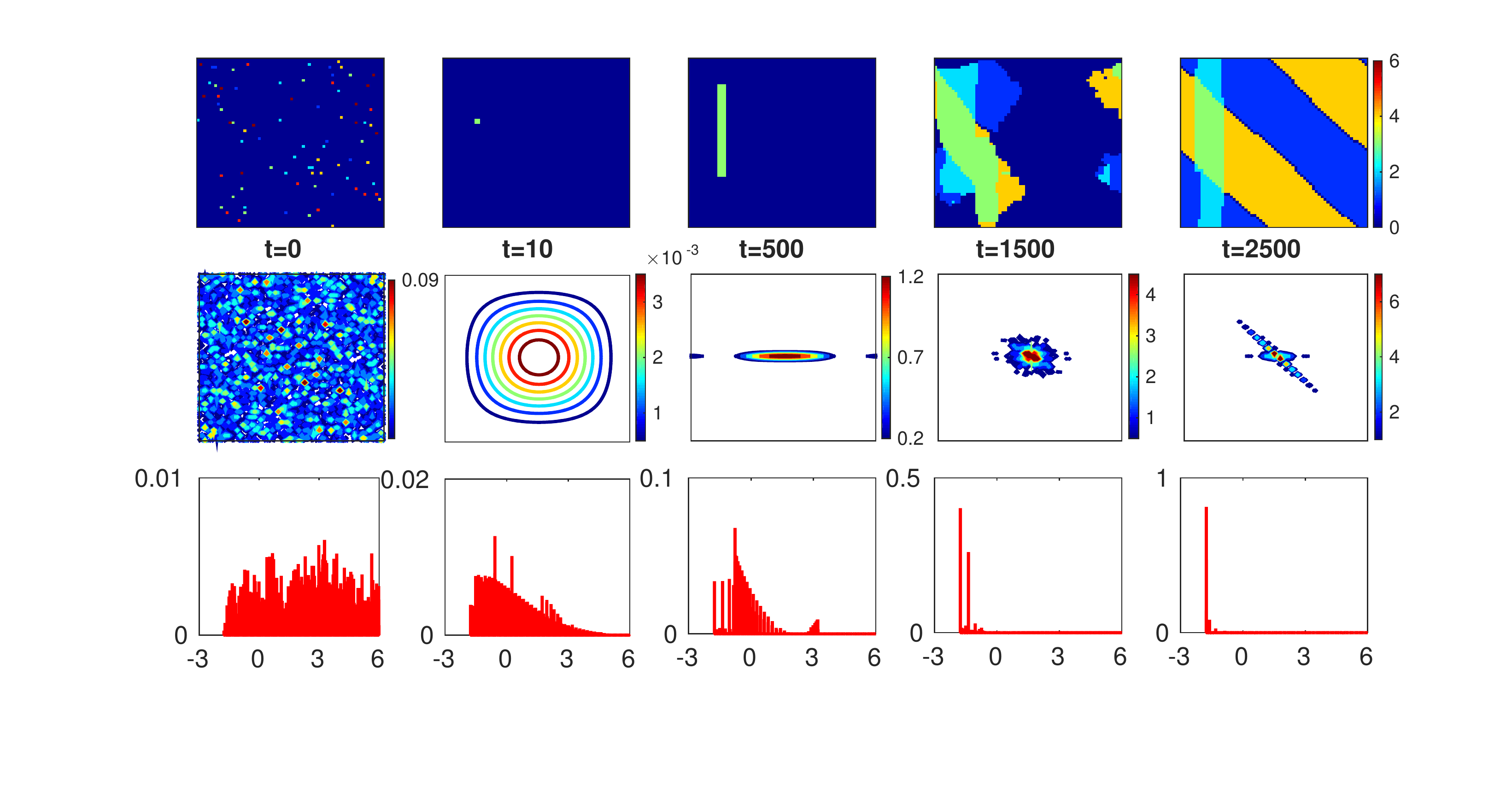}
\caption{(Color online) {\it Evolution of strain textures and energy occupancy in TO transition:} On quench-and-hold to $T=0.79$, snapshots of OP textures in Brillouin zone as $ln(1+|\vec S(\vec k)|^2)$ contours ({\it second row}) of corresponding coordinate space textures ({\it first row}) showing incubation during domain-wall vapour to liquid which finally converts to crystal. See movies online \cite{R39} of these evolutions in both coordinate and Fourier spaces. The energy occupancy $\rho_{22}(\epsilon,t)$ of structure factor vs $\epsilon_{22}(\vec k)$ ({\it third row}) showing the shifting of density of states when the distribution enters into the golf-course. See text. }
\end{center}
\label{Fig.10}
\end{figure*}

\section{Results and Discussions}
 
\subsection{Golf-courses, funnels, and conversion times}

On quenching to a temperature $T_2<T<T_4$, figure 3 shows the singlerun Hamiltonian energy of (2.8a) that is quite flat at $H(t) \equiv H_{22}+H_{23}+H_{33} \simeq 0 $ (and $H_{22}=H_{23}=H_{33} \simeq 0$) during incubation and then falls rapidly to lower energies in TCR, SO, and TO transitions. So the relavent spectrum is thus a zero-energy plane through the $\epsilon_{\ell\ell^{'}}(\vec k)=0$ or $\epsilon_{22}(\vec k)=\epsilon_{23}(\vec k)=\epsilon_{33}(\vec k) \simeq 0$ surface and the negative energies below it. The resulting relief plot of $\epsilon_{\ell \ell^{\prime}}(\vec k)$ is shown in Figure 4 (top row) that depicts a momentum space anisotropic golf-course defined by $\epsilon_{\ell \ell^{'}}(k)=0$ and funnel for $\epsilon_{\ell \ell^{'}}(k)<0$ inside it. 
Figure 4 (bottom row) shows the temperature-dependent, anisotropic golf-course edge that is large (small) at low (high) temperatures. Such energy landscape concepts are used in configuration space to study the rapid and slow folding of proteins \cite{R2,R3,R4,R8}.  

Figure 5 shows the singlerun martensite fraction $n_m (t)$ versus time $t$ after quenches to $T_1, T_2$ and $T_4$ that have different values in different transitions \cite{R21}. See Figure 2. The martensite fraction $n_m (t)$ rises rapidly to unity and conversion times $t_m \simeq 1$ MCS at low temperatures $T=T_1$ where the golf-course edge is large. At moderate temperatures $T=T_2$, the golf-course edge is moderate and $t_m \simeq 10$ MCS. As the transition is approached  $T \simeq T_4(<<T_0)$, $n_m(t)$ shows {\it  incubation} behaviour and $t_m \sim 10^3-10^4$ MCS where the golf-course edge is small \cite{R26}. 
For $T>T_4$, the four petaled golf-course topology provides an infinite entropy barrier for conversions \cite{R26}.

The temperature dependence of conversion times \cite{R21} ${\bar t_{m}}(T)$ and conversion success-fraction $\phi_m$ for a fixed elastic stiffness $A_1$ and different Hamiltonian energy scales $E_0$ is plotted in Fig. 6 for TCR, SO, and TO transitions. 
The conversion success-fraction $\phi_m$ is unity for $T<T_1$, where conversion times $\bar t_m \sim 1$ MCS, and decreases linearly at $T=T_2$, where $\bar t_m \simeq 10$ MCS, to become $\phi_m=0$ at $T \simeq T_4$ with Vogel-Fulcher conversion times \cite{R10}  ${\bar t}_{m} = t_0 \exp [ b_0 |T_1 - T_4|/| T- T_4|]$, with $t_0 = 1.6, b_0 = 1.7$. The success-fraction $\phi_m$ and conversion times $\bar t_m$ found insensitive to energy scales $E_0$ and hence are understood to arise from the dominant entropy barriers that vanish at $T=T_1$ and diverge at $T \simeq T_4$ with vanishing of rare conversion pathways \cite{R17}.

We calculate the arithmetic mean rate $< r_m> \equiv <1/t_m>$ that determines ${\bar t}_m= 1/<{\bar r_m}>$, with $1/t_h < r_m < 1$ in TCR, SO, and TO transitions. The variance in the rates  is ${\sigma^{2}_{r_{m}}} = <( r_m -< r_m>)^2>$. The probability densities $P(r_m)$ versus $r_m$ for various $\Delta T$ are shown in Fig 7, as histograms for different temperatures. For each histogram of $N_{hist}$ data points, the Scott optimized bin size \cite{R38}  is used, of $dr_m = 3.5 \sigma_r /[N_{hist}]^{1/3}$, as in the SR case \cite{R17}. The histograms again narrow sharply for $T<T_2$, as in the delta-function-like  peak on the right. See also Figs. 5 and 6. For calculated $<r_{m}>$ and $\sigma_{{r}_{m}}^2$ from the data, the best fits shown as solid lines are the Log-normal curves that are signatures of rare-events \cite{R33}.

\subsection{Evolution of strain textures in Fourier space}

For deep quenches $T<T_1$, the edge is large and hence the structure factor distribution $ln[1+|\vec S(\vec k,t)|^2]$ rolls into the golf-course quickly within $\bar t_m \sim 1$ MCS. For moderate quenches $T_1<T<T_2$, the edge is also moderate and hence the distribution enters into the golf-course in $\bar t_m \sim 10$ MCS. To study the re-equilibration and nature of the entropy barriers, we consider the shallow quenches $T_2<T<T_4$ where the distribution shows {\it ageing} or {\it incubation} to enter into the golf-course. 

After a temperature quench, we track the strain-pseudospin $\vec S=(S_2,S_3)$ textures in terms of variant label $V$ that can be $V=0$ in the austenite and $V=1,2,..,N_{v}$ in the martensite to represent the variants of (2.8c) for TCR, SO, and TO transitions.
We define energy occupancy $\rho(\epsilon,t)$ or Fourier intensity at a given $\vec k$, as in protein folding simulations \cite{R9},

$$\rho(\epsilon, t) = \frac{\sum_{\vec k} \sum_{\ell,\ell^{\prime}=2,3} \delta_{ \epsilon_{\ell\ell^{\prime}}, \epsilon_{\ell\ell^{\prime}} (\vec k)}   \rho(\vec k, t)}{ \sum_{\vec k} \rho(\vec k, t)}.~~(3.1)$$

The Figs. 8, 9, and 10 shows the singlerun evolution of the strain textures both in coordinate and Fourier spaces and also the energy occupancy for TCR, SO and TO transitions. See movies online \cite{R39} of these evolutions in Supplemental material. On quenching the dilutely seeded austenite into $T_2<T<T_4 (<<T_0)$, the coordinate space textures (first row) as found earlier in \cite{R21} shows that the initial $t=0$ dilute martensite seeds in the austenite disappear quickly to form single variant droplet(s), reminding of Ostwald ripening, to form DW vapour. The incubating vapour droplet(s) grows through fluctuations and {\it autocatalytic twinning} or {\it elastic photocopying} \cite{R34,R35} to convert to DW liquid of wandering walls. The domain walls then orient at a later time into the preferred crystallographic directions to form DW crystal. 

The conversion-incubation time is best understood in Fourier space. The second row of main Figs. 8, 9, and 10 shows the same evolving strain  textures but now in Fourier space as contour plots. 
The initial distribution of dilute martensite seeds at $t=0$ rapidly convert to isotropic gaussian distribution (broad $+$-shape distribution in case of TCR) of DW vapour that incubates and generates wings along the $k_x$-axis to reduce in size with an increase in height as in the SR case \cite{R17}. 
The wings along $k_x$-axis (both axes in case of TCR) persists for long time before generating wings along $k_y$-axis during elastic photocopying.
The anisotropies along both axes reduces and width becomes small in size for the $+$-shape distribution of DW liquid to fit and enter into the golf-course at $t=t_m$. 
Finding out these constant-energy anisotropic pathways constitute an entropy barrier. Once inside the funnel, the distribution of liquid orients along the preferred directions to form the DW crystal.

The evolution of energy occupancy $\rho_{22}(\epsilon,t)$ versus $\epsilon_{22}(\vec k)$ is shown in the third row of main Figs. 8, 9, and 10. The evolution of the total occupancy $\rho(\epsilon,t)$ versus $\epsilon$ shows the similar behaviour (not shown). The edge of the golf-course is $\epsilon_{22}(\vec k)=0$. In the vapour phase and during the incubation, the occupancy is small and remains same. When the wings are generated, a small peak is seen at higher energies in the occupancy.  At $t=t_m$, when the entropy barrier is crossed, the distribution enters into the golf-course and the occupancy moves into the negative ($g_L(T)<\epsilon(T)<0$) energy funnel.
In TCR, SO, and TO transitions, the final 'equilibrium' distribution (not shown) is an inverse-energy falloff in the excitation energy above the bulk Landau term, $\tilde \epsilon \equiv \epsilon - g_L > 0$ as in the SR case \cite{R26},
 $$\rho( {\tilde \epsilon}, t; T) \rightarrow 1/{\tilde \epsilon}, ~~~(3.2)$$
 that is found in inhomogeneous harmonic oscillators \cite{R13}.

\subsection{Textural thermodynamics and acceptance fractions}

At a given Monte Carlo sweep $t$, the expressions for the free energy $F \simeq F_{LMF} (t)$, internal energy $U (t)$, and entropy  $S_{entr} (t)$, in terms of the $\{ \vec S(\vec r,t)\}$ configurations, are obtained \cite{R40} from partition functions \cite{R22} for vector-OP TCR, SO, and TO transitions following the same procedure as used in scalar-OP SR transition \cite{R17,R20,R35}.

 \begin{figure*}[ht]
\begin{center}
\includegraphics[height=4.3cm, width=16.0cm,trim=0.0cm 0.0cm 0.05cm 0.0cm,clip]{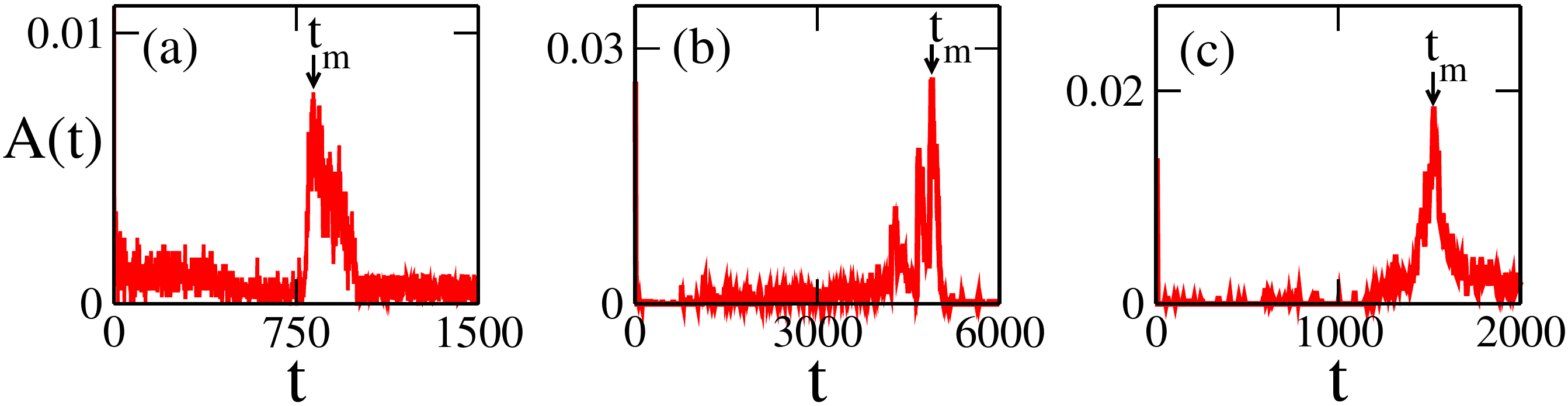}\\
\includegraphics[height=4.3cm, width=16.0cm,trim=-0.5cm 0.0cm 0.05cm 0.0cm,clip]{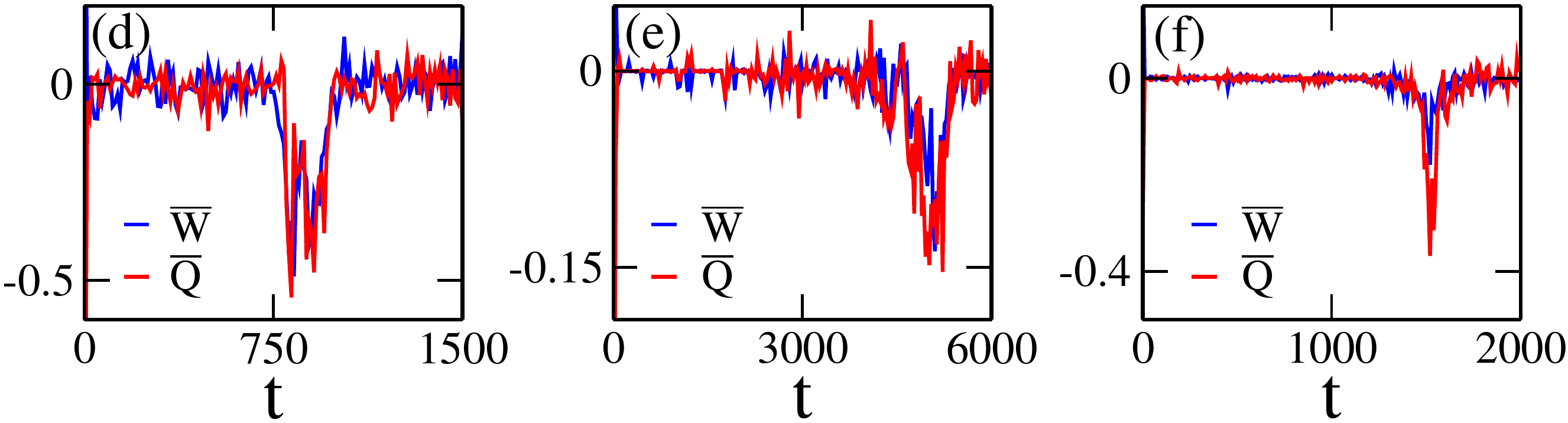}
\caption{(Color online) {\it Monte Carlo acceptance fractions and domain-wall thermodynamics:} The Monte Carlo acceptance fraction $A_{act}(t)$ versus time $t$ is almost zero in the incubating state and shows a peak at martensite conversion time $t=t_m$. The rates of work and heat releases $\dot W$ and $\dot Q$ are also zero in the incubating state and shows a dip with large releases at $t=t_m$. In the 'equilibrium', both the acceptance fractions and rates of heat and work releases are again zero as shown in (a,d) TCR (b,e) SO and (c,f) TO transitions. See text.}
\end{center}
\label{Fig.11}
\end{figure*}

After a temperature quench, the total change in the internal energy is, by a First-Law of thermodynamics-type relation,
$$ dU (t) = \dbar W(t) + \dbar Q(t), ~~~(3.3)$$
where ${\dbar}W(t)=dF_{LMF}(t)$ is the work done by the domain walls and $\dbar Q (t)=TdS_{entr}(t)$ is the heat release by the spins at bath temperature \cite{R26}. 
One can track the relative changes of $\dbar Q, \dbar W$ through an effective temperature $T_{eff}$ from $\dbar W(t)=[1-\frac{T}{T_{eff}}]  dU (t)$ and $\dbar Q= \frac{T}{T_{eff}} dU(t)$ that is similar to the 'microcanonical' definition,
$$ \frac{T}{T_{eff}(t)}  = T \frac{d S (t)}{ d U (t)}, ~~~(3.4)$$
as used in protein folding models \cite{R9}. The effective temperature reaches the bath temperature in equilibrium $T_{eff}(t) \rightarrow T $, where the local internal stresses vanish \cite{R26}. The detailed study of the DW liquid to DW crystal and effective temperature will be pursued elsewhere.

The single run rates of heat and work emissions by the domain walls are shown in Figure 11 (d),(e),(f), where the rates are $\dot X=X(t+1)-X(t)$. The rates are zero in the ageing state and large at $t=t_m$ where the entropy barrier is crossed \cite{R26}. The rates again become zero in the equilibrium.
The MC acceptance fractions $A_{act}(t)$ are shown in Fig. 11 (a),(b),(c) for TCR, SO, and TO transitions respectively. Notice, $A_{act}(t)$ are roughly zero during incubation and rises to peak at $t=t_m$ to signal crossing of the entropy barrier \cite{R26} and becomes zero again in the equilibrium. 

On cooling, the incubation for transition and transition enthalphy and entropy can be calculated \cite{R41,R42,R43,R44,R45,R46} systematically for martensitic transitions in 2- and 3-spatial dimensions, which could be pursured in our further study \cite{R47}.

\section{Summary and further work}

We do systematic temperature-quench Monte Carlo simulations to study the re-equilibration in the athermal martensites using protein folding concepts such as golf-courses and funnels {\it that appear naturally}, in our vector-OP ($N_{OP}=2$) 4-state ($N_{v}+1=4$), 5-state ($N_{v}+1=5$) and 7-state ($N_{v}+1=7$) strain-pseudospin clock-zero model Hamiltonians for triangle-centered rectangle, square-oblique and triangle-oblique transitions.
The simulation results are as follows: (i) The energy landscape concepts such as golf-courses and funnels of protein folding and Monte Carlo acceptance fractions from harmonic oscillators turn out to be very useful in understanding the re-equilibration process in athermal martensites.
The incubation-delays and rapid conversions in the temperature-time-transfomation phase diagram are understood from the presence of small and large edge of the golf-course respectively.
(ii) The incubation-delay times that are insensitive to Hamiltonian energy scales are found to have Log-normal distributions, which are signatures of rare events. The conversion success-fraction also found insensitive to energy scales becomes zero at the Vogel-Fulcher transition temperature with diverging entropy barriers from vanishing of rare pathways. 
(iii) The DW vapor to liquid conversion-incubation in coordinate space is understood best in Fourier space as the ageing for the structure factor distribution to find constant-energy anisotropic pathways while facing entropy barrier to enter into the golf-course. This is reflected in the occupancy as shifting of density of states into the negative funnel region. Once inside the funnel, the distribution of DW liquid orient later to form DW crystal. 
(iv) Monte Carlo acceptance fractions show a peak and heat and work releases show a dip when the entropy barrier is crossed, which are zero in the ageing state. An effective temperature can be defined similar to protein folding models that reaches bath temperature in the equilibrium when local internal stresses vanish.



Further work could also include systematic MC temperature quench simulations to study the re-equilibration using protein folding concepts in strain-pseudospin clock-zero model Hamiltonians for athermal martensitic transitions in 3-spatial dimensions \cite{R47}.

{\it Acknowledgements:} It is a pleasure to thank Subodh R. Shenoy, K.P.N. Murthy, T. Lookman, Sanjay Puri and Surajit Sengupta for useful discussions, encouragement and support. Part of this work is done during UGC- Dr. D.S. Kothari Postdoctoral Fellowship.

\end{document}